\documentstyle[twocolumn,epsf]{jpsj}
\newcommand{\citePRSA}[3]{Proc.\ Roy.\ Soc.\ A {\bf #1} (#2) #3}
\newcommand{\citePR}[3]{Phys.\ Rev. {\bf #1} (#2) #3}
\newcommand{\citePRB}[3]{Phys.\ Rev.\ B {\bf #1} (#2) #3}
\newcommand{\citePRL}[3]{Phys.\ Rev.\ Lett. {\bf #1} (#2) #3}
\newcommand{\citeRMP}[3]{Rev.\ Mod.\ Phys. {\bf #1} (#2) #3}
\newcommand{\citeJPSJ}[3]{J.\ Phys.\ Soc.\ Jpn. {\bf #1} (#2) #3}

\newcommand{\citePhysicaC}[3]{Physica\ C\ {\bf #1} (#2) #3}

\newcommand{\citeIBID}[3]{ibid. {\bf #1} (#2) #3}

\makeatletter
\def\lsim{\mathrel{\mathpalette\glsim@align<}}
\def\gsim{\mathrel{\mathpalette\glsim@align>}}
\def\glsim@align#1#2{\lower.6ex\vbox{\baselineskip\z@skip\lineskip\z@\ialign{$\m@th#1\hfil##\hfil$\crcr#2\crcr\sim\crcr}}}
\makeatother

\def\runtitle{Mott Transition v.s. Multicritical Phenomenon
of Superconductivity and Antiferromagnetism 
-- Application to $\kappa$-(BEDT-TTF)$_2$X --}

\def\runauthor{Shigeki \textsc{Onoda} and Naoto \textsc{Nagaosa}}

\title{Mott Transition vs Multicritical Phenomenon
of Superconductivity and Antiferromagnetism
-- Application to $\kappa$-(BEDT-TTF)$_2$X --}

\author{Shigeki {\sc Onoda}\thanks{E-mail address: sonoda@appi.t.u-tokyo.ac.jp}
 and Naoto \sc{Nagaosa}$^{1,2,3}$
}

\inst{Tokura Spin Superstructure Project, ERATO, Japan Science and Technology Corporation, c/o Department of Applied Physics, University of Tokyo, Hongo 7-3-1, Tokyo 113-8656, Japan\\
$^1$Department of Applied Physics, University of Tokyo, Hongo 7-3-1, Tokyo 113-8656, Japan\\
$^2$Correlated Electron Research Center, AIST, Tsukuba, Ibaraki 305-0046, 
Japan\\
$^3$CREST, Japan Science and Technology Corporation}

\recdate{\today}
\abst{
Interplay between Mott transition and the multicritical phenomenon of 
$d$-wave superconductivity (SC) and antiferromagnetism (AF) is studied 
theoretically. We describe the Mott transition, which is 
analogous to a liquid-gas phase transition, in terms of an Ising-type order 
parameter $\eta$. We reveal possible mean-field phase diagrams produced by 
this interplay. Renormalization group analysis up to one-loop order gives 
flows of coupling constants, which in most cases lead to fluctuation-induced 
first-order phase transitions even when the SO(5) symmetry exists betwen the 
SC and AF. Behaviors of various physical quantities around the Mott 
critical point are predicted. Experiments on $\kappa$-(BEDT-TTF)$_2$X are 
discussed from this viewpoint.}
\kword{organic superconductivity, Mott transition, antiferromagnetism, renormalization group}

\begin{document}
\sloppy
\maketitle


Emergence of magnetism and superconductivity (SC) around Mott transition 
\cite{Mott} has been one of the central issues in strongly correlated electron 
systems. The Mott transition \cite{Mott} is referred to as a phase transition driven by a local Coulomb repulsion from metal to Mott insulator, i.e., an insulator with local spin moment, whether it possesses an antiferromagnetic long-range order (AFLRO) or not. There are two distinct pictures for the Mott-insulating states: 
(I) In Slater's picture \cite{Slater} justified in the weak correlation limit, 
the insulating state originates from doubling of a unit cell 
accompanied by the AFLRO. 
(II) In the Mott-Hubbard picture \cite{Mott,Hubbard} valid in the strong 
correlation limit, the system can be insulating without any symmetry 
breaking. This insulating feature arises from the Hubbard gap for the
charge transfer due to a strong local Coulomb repulsion $U$. 
The superexchange interaction between localized spins causes
an antiferromagnetic phase transition at a temperature much lower
than the Hubbard gap or $U$. 
Therefore, Mott transitions should be dealt 
with as a different degrees of freedom from antiferromagnetism (AF).

The dynamical mean field theory (DMFT) has revealed that the Mott transition 
without any symmetry breaking occurs for the Hubbard model in infinite 
dimensions \cite{DMFT}. A first-order Mott transition curve appears by varying 
a ratio of bandwidth $W$ to $U$ and terminates at a finite-temperature 
critical end point (Mott critical point). This Mott transition is analogous to 
the liquid-gas phase transition, as pointed out in a field theory \cite{Castellani79}. Recently, by restoring spatial correlations neglected in the DMFT, 
the correlator projection method has given a similar structure of 
Mott transition in two dimensions but modified by 
spin correlations \cite{OnodaImada03lett}. 
Diverging compressibility is a common feature at the Mott critical point 
\cite{KotliarMurthyRozenberg02,OnodaImada01full}.

To describe Mott transitions, one can introduce an Ising-type ``order parameter'' $\eta$ with its spatial variations as in the case of a lattice gas model. Here, we note that $\eta$ is not associated with any symmetry breaking. Actually, Kotliar {\it et al.} have developed Ginzburg-Landau (GL) theory of Mott transition in infinite dimensions where spatial correlations are absent \cite{KotliarLangeRozenberg00}. For finite-temperature Mott transitions, the order parameter $\eta$ describes a static eigen mode of the dynamical mean field. Then, to discuss 
multicritical phnomena of the AF and the SC around the Mott transition, 
we need to introduce a scalar field $\eta$, an 
O(2) vector field $\vec{\Delta}$ for the SC and an 
O(3) vector field $\vec{m}$ for the AF. 

\begin{figure}[htb]
\begin{center}
\epsfxsize=7.0cm
$$\epsffile{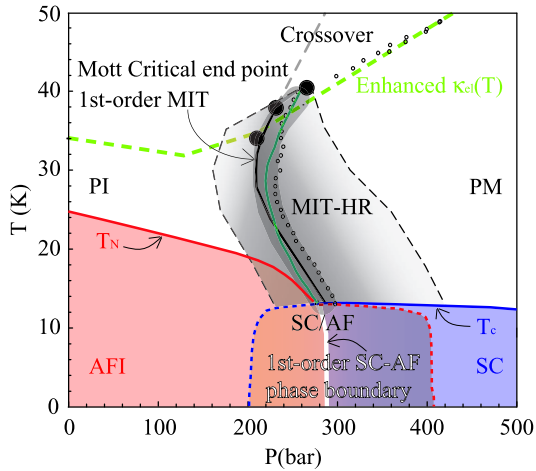}$$
\end{center}
\vspace*{-5mm}
\caption{(Color) Experimental phase diagram of $\kappa$-(BEDT-TTF)$_2$Cu[N(CN)$_2$]Cl with paramagnetic metal (PM), paramagnetic insulator (PI), antiferromagnetic insulator (AFI), inhomogeneously coexistent region of AF and SC (SC/AF), and hysteresis region of metal-insulator transition (MIT-HR). $T_{\rm c}$ is determined from AC susceptibility $\chi'$ \cite{Lefebvre00} and anomaly in ultrasonic velocity $c$ \cite{Fournier03}, while $T_{\rm N}$ from NMR \cite{Lefebvre00}. The first-order SC-AF phase boundary is determined from these three probes \cite{Lefebvre00,Fournier03}. Dotted curves represent spinodal instabilities for metastable AF (red) and SC (blue) determined from NMR \cite{Lefebvre00}, $\chi'$ \cite{Lefebvre00}, and anomaly in $c$ \cite{Fournier03}.
The first-order MIT curve is determined from $\chi'$ (green solid curve) \cite{Lefebvre00}, jump in resistivity $\rho$ (black solid curve) \cite{Kanoda_unpublished} and maximum in $d\rho/dP$ (open circles) \cite{Limelette03}, all of which lie in the dark shaded region. The MIT-HR is surrounded by spinodal instabilities (black broken curves) determined from hysteresis in $\rho$ \cite{Limelette03}. The Mott critical point is the end point (black filled circles for different experiments \cite{Lefebvre00,Fournier03,Limelette03,Kanoda_unpublished}) of the observed first-order MIT curves. Crossover signaled by enhanced charge compressibility (green dashed curve) is determined from dip in $c$ \cite{Fournier03}.}
\label{fig:exp}
\end{figure}
Layered organic superconductor $\kappa$-(BEDT-TTF)$_2$X 
offers an ideal laboratory to address this issue.
It exhibits a Mott transition as the effective ratio $U/W$ is varied by 
changing X, isotope effect, and applying pressure \cite{Ito96,Kanoda97}. 
A phase diagram determined by recent experiments on the Cl salt (X=Cu[N(CN)$_2$]Cl) is 
summarized in Fig.~\ref{fig:exp}: There exists a first-order Mott transition 
curve separating SC-AF phases at low temperature, which 
continues to higher temperature terminating at the Mott critical point
$(P_{\rm cr},T_{\rm cr}) \cong (260\ {\rm bar}, 40\ {\rm K})$ 
\cite{Lefebvre00,Fournier03,Limelette03,Kanoda_unpublished,Kanoda97}. 
The Mott critical point has been observed with an 
anomaly in sound velocity as diverging charge compressibility 
$\kappa_{\rm el}$ \cite{Fournier03} which has been theoretically predicted 
\cite{DMFT,KotliarMurthyRozenberg02,OnodaImada01full}. 
This Mott transition appears with a similarity 
to a liquid-gas phase transition, as in other systems \cite{Mott} 
and theoretical results \cite{Castellani79,DMFT,OnodaImada03lett}.

In this compound, the Neel temperature $T_{\rm N}$ and 
the superconducting critical temperature $T_{\rm c}$ meet at
$(P_{\rm bcr},T_{\rm bcr}) \cong (280\ {\rm bar}, 13\ {\rm K})$, 
showing the features of a bicritical phenomenon 
\cite{Ito96,Kanoda97,Lefebvre00} (Fig. 1). 
GL theory for these two competing orders has been 
proposed \cite{MurakamiNagaosa00}, and the analysis of the dynamical critical 
properties of NMR $T_1^{-1}$ revealed the SO(5) symmetry 
between the SC and AF, which was first proposed for high-Tc cuprates
\cite{Zhang97}. The SO(5)-symmetric fixed point yields bicritical phenomena 
where order-disorder phase transitions become second-order. In contrast, away 
from the SO(5) symmetry, a fluctuation-induced first-order phase transition 
from disordered to ordered phase \cite{fluctuation-induced-1st-order} 
or a coexistent phase appears, depending on the strength of competition.

Now, it is highly important to clarify the interplay between Mott transition 
and the competing two orders, i.e., AF and SC. In this paper, we develop a 
GL theory for this purpose. Global phase diagram is studied
including the fluctuation up to one-loop order. Critical properties around the 
Mott critical point 
are also discussed. 

We start from the following free energy expansion in terms of three order parameters $\eta$, $\vec{\Delta}$ and $\vec{m}$;
\begin{eqnarray*}
F&=&\frac{1}{2}\int\!d{\mib r}\Big[r_\eta\eta({\mib r})^2+(\mib \nabla\eta({\mib r}))^2+\frac{u_{\eta\eta}}{2}\eta({\mib r})^4-2h\eta({\mib r})
\nonumber\\
&&+r_\Delta\vec{\Delta}({\mib r})^2+(\mib \nabla\vec{\Delta}({\mib r}))^2+\frac{u_{\Delta\Delta}}{2}(\vec{\Delta}({\mib r})^2)^2
\nonumber\\
&&+r_m\vec{m}({\mib r})^2+(\mib \nabla\vec{m}({\mib r}))^2
+\frac{u_{mm}}{2}(\vec{m}({\mib r})^2)^2
\nonumber\\
&&+u_{\Delta m}\vec{m}({\mib r})^2\vec{\Delta}({\mib r})^2
-(g_\Delta\vec{\Delta}({\mib r})^2-g_m\vec{m}({\mib r})^2)\eta({\mib r})\Big].
\end{eqnarray*}
$h$ plays a crucial role in describing the Mott critical point 
as introduced by Kotliar {\it et al.} in the original arguments of 
GL theory of Mott transition \cite{KotliarMurthyRozenberg02}. 
It is expected to obey $h\sim h_0 (P-P_0(T))$ around the Mott critical point 
where $P$ controls $(U/W)^{-1}$ and $P_0(T)$ represents its value at the metal-insulator phase boundary. $r_i$ represents a distance from the bare critical point at temperature $T_{i0}$ ($i=\eta$, $\Delta$, $m$). It has the form, $r_i\sim a_i(T-T_{i0}(P))$. For SC and AF degrees of freedom, we take into account the effects of $P$ on $T_{i0}$ by expanding it around the Mott critical 
 point as $T_{i0}(P)\sim T_{i0}(P_0(T_{\eta0}))+b_i(P-P_0(T_{\eta0}))$ with $b_\Delta>0$ and $b_m<0$. 
$u_{ij}$ ($i,j=\eta$, $\Delta$, $m$) describes a quartic coupling constant. $g_\Delta$ ($g_m$) is a lowest-order coupling constant between $\eta$ and $\vec{\Delta}$ ($\vec{m}$).

\begin{figure}[tb]
\begin{center}
\epsfxsize=6.0cm
$$\epsffile{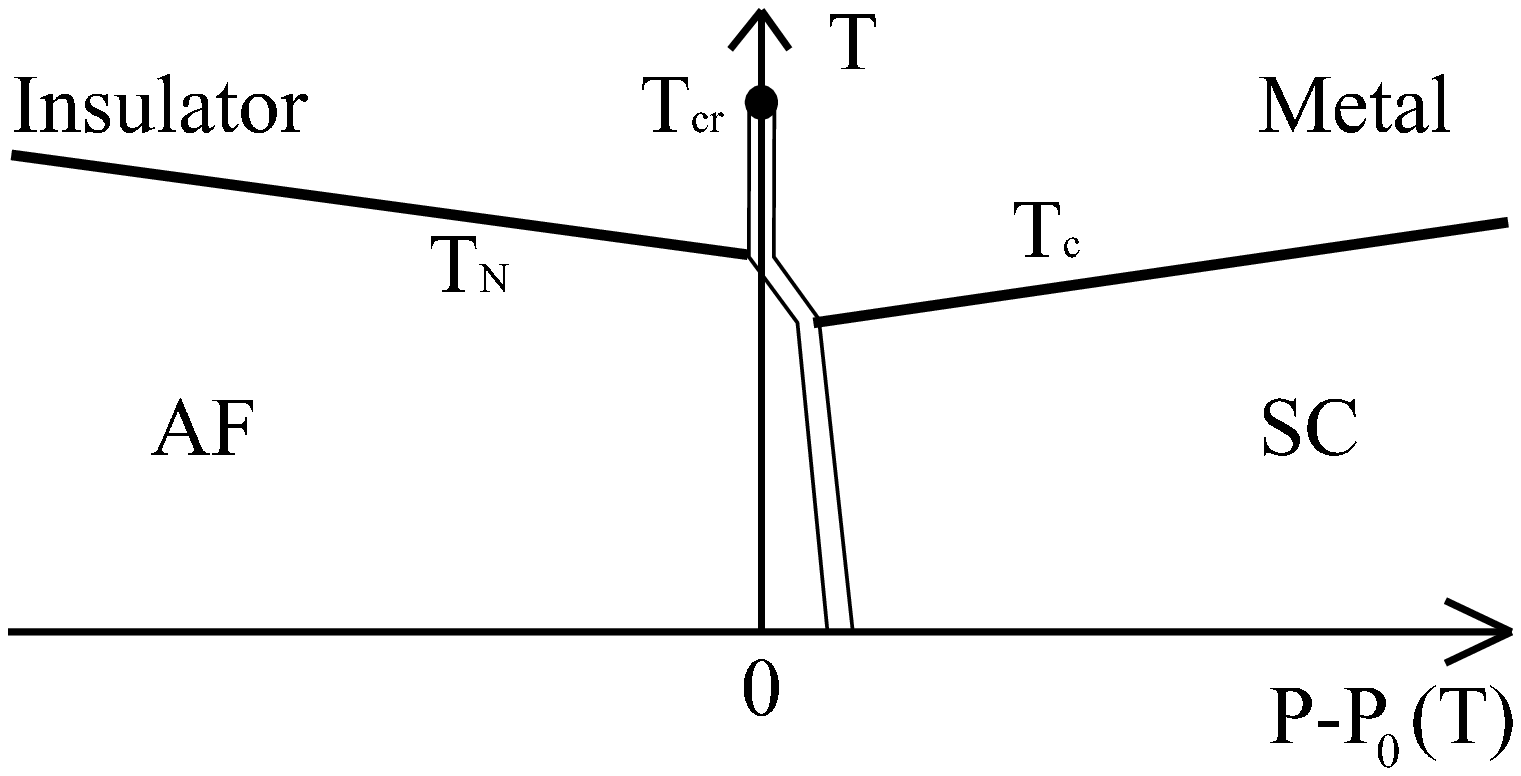}$$
\end{center}
\vspace*{-12mm}
(a)
\begin{center}
\epsfxsize=6.0cm
$$\epsffile{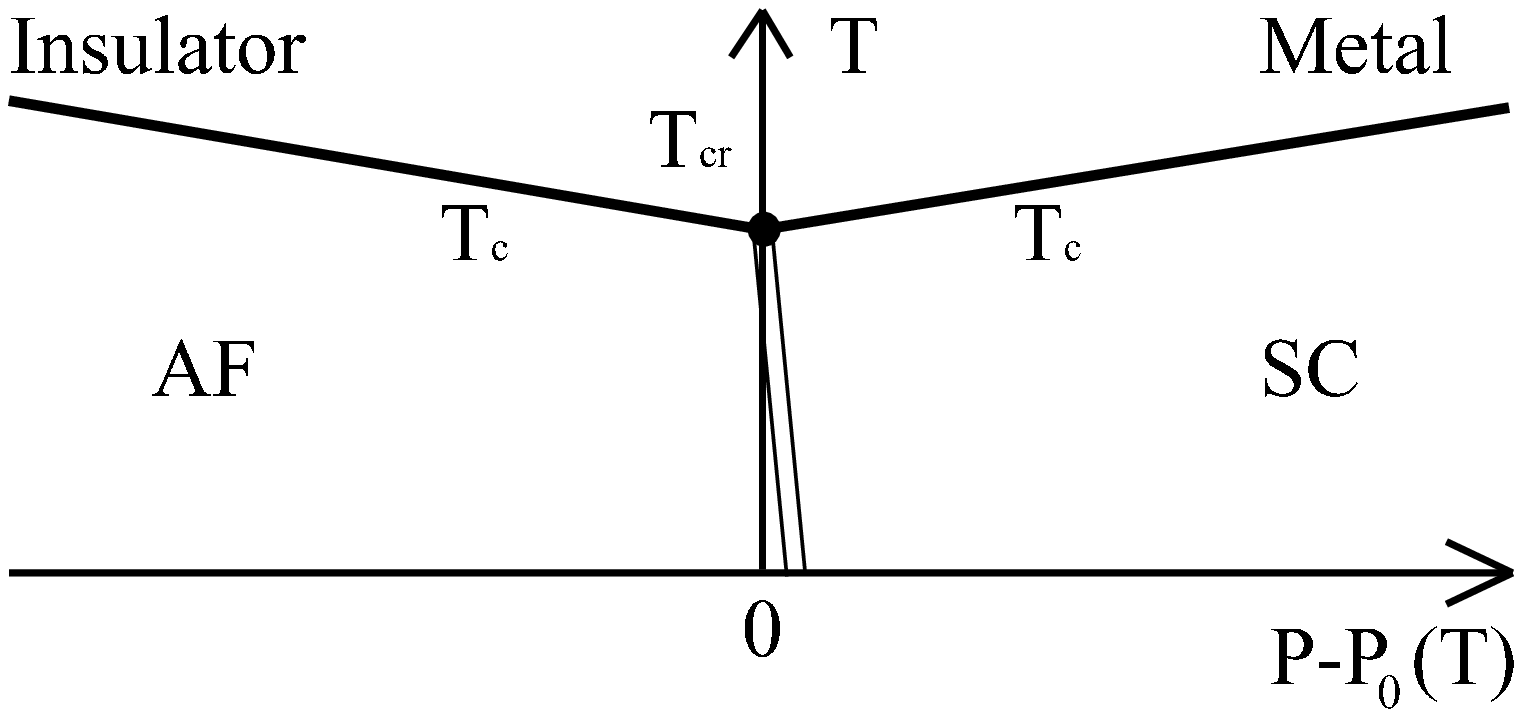}$$
\end{center}
\vspace*{-22mm}
(b)
\begin{center}
\epsfxsize=6.0cm
$$\epsffile{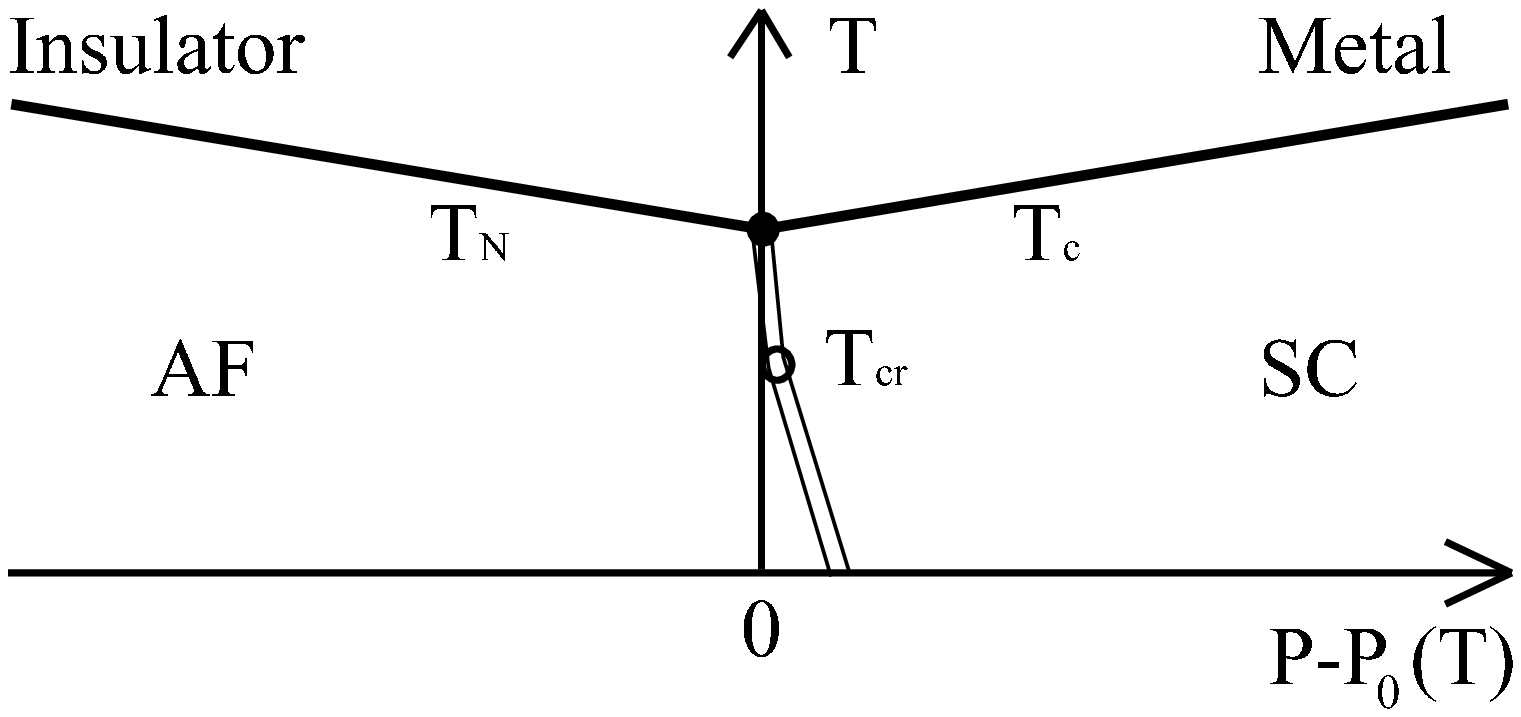}$$
\end{center}
\vspace*{-12mm}
(c)
\caption{Schematic MF phase diagrams. Bold and doubled lines represent second-order and first-order transitions, respectively, while circles the Mott critical point.}
\label{fig:MF}
\end{figure}

First, we study the mean-field (MF) phase diagrams. If $u_{\Delta m} \gg u_{\Delta\Delta},u_{mm}$, the coexistence of SC and AF is prohibited. We also restrict ourselves to the case that at $T<T_{\rm cr},T_{\rm c},T_{\rm N}$, the first-order Mott transition curve separates the SC and the AF with $T_{\rm N} \ge T_{\rm c}$, which is singnificantly important related to the $\kappa$-(BEDT-TTF)$_2$X system.
Then, there are three possibilities as shown in Fig.~\ref{fig:MF}: In (a), the first-order Mott transition runs out above $T_{\rm C}$ and $T_{\rm N}$ and ends at a critical point. If $T_{\rm cr} \gg T_{\rm c}, T_{\rm N}$, the average $\langle \eta \rangle$ takes $(-)\sqrt{|r_\eta|/u_{\eta\eta}}$ in the metallic (insulating) side. Then, an effective value for $r_\Delta-r_m$ is modified by $(g_\Delta+g_m)\langle \eta \rangle$, which exhibits a jump given by $\delta r_{\Delta - m} =2(g_\Delta+g_m)\sqrt{|r_\eta|/u_{\eta\eta}}$ at the Mott transition curve. In (b), the bicritical point of SC-AF transition coincides with the Mott critical point. Destiniy of this phase diagram is discussed below using a renormalization group (RG) analysis. In (c), the Mott critical point is inside the SC-AF phase boundary and $\eta$ is massive at the bicritical point. Then, the competition between AF and SC becomes the main issue \cite{Zhang97,MurakamiNagaosa00}.

\begin{figure}[tb]
\begin{center}
\epsfxsize=6.0cm
$$\epsffile{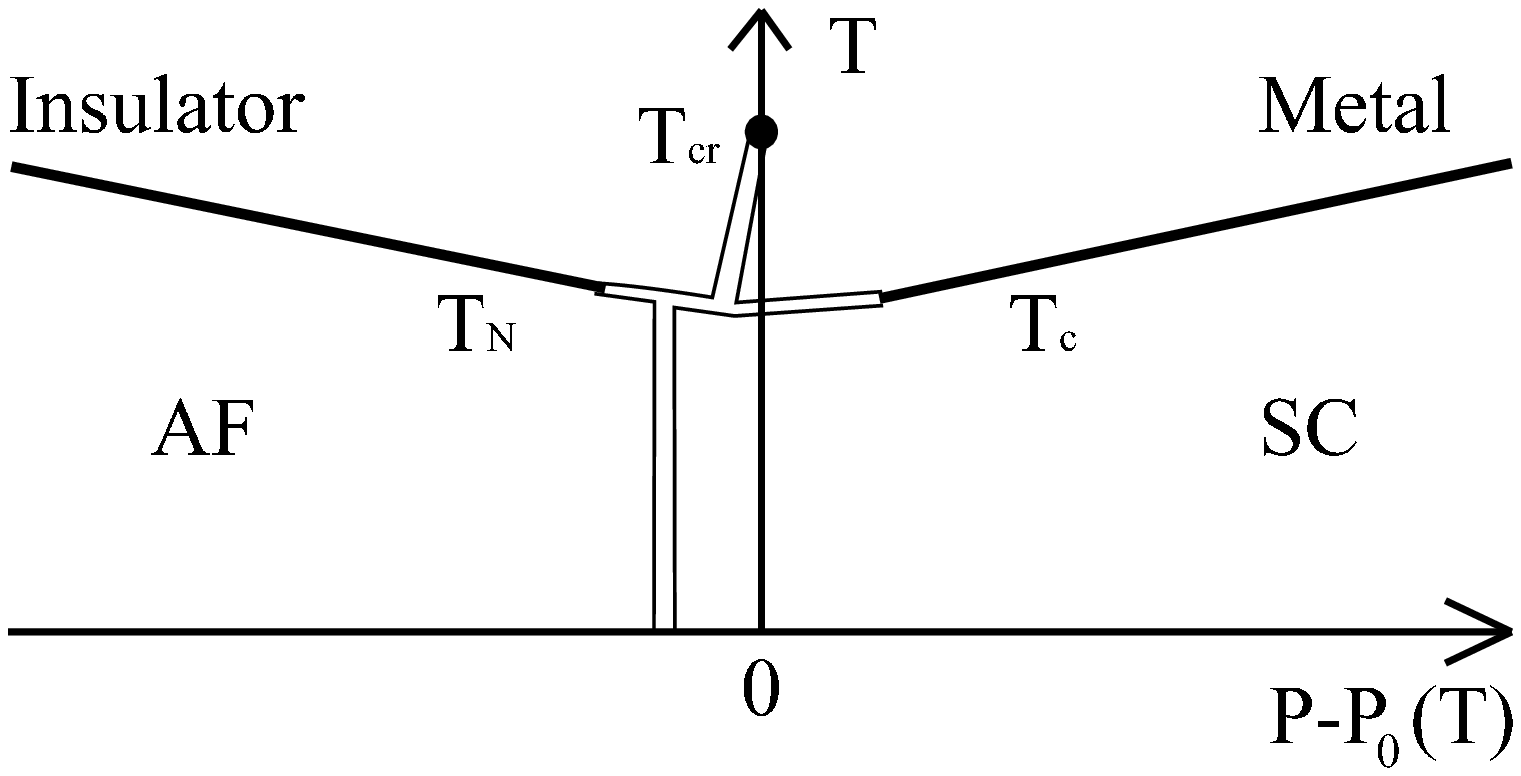}$$
\end{center}
\vspace*{-12mm}
(a)
\begin{center}
\epsfxsize=6.0cm
$$\epsffile{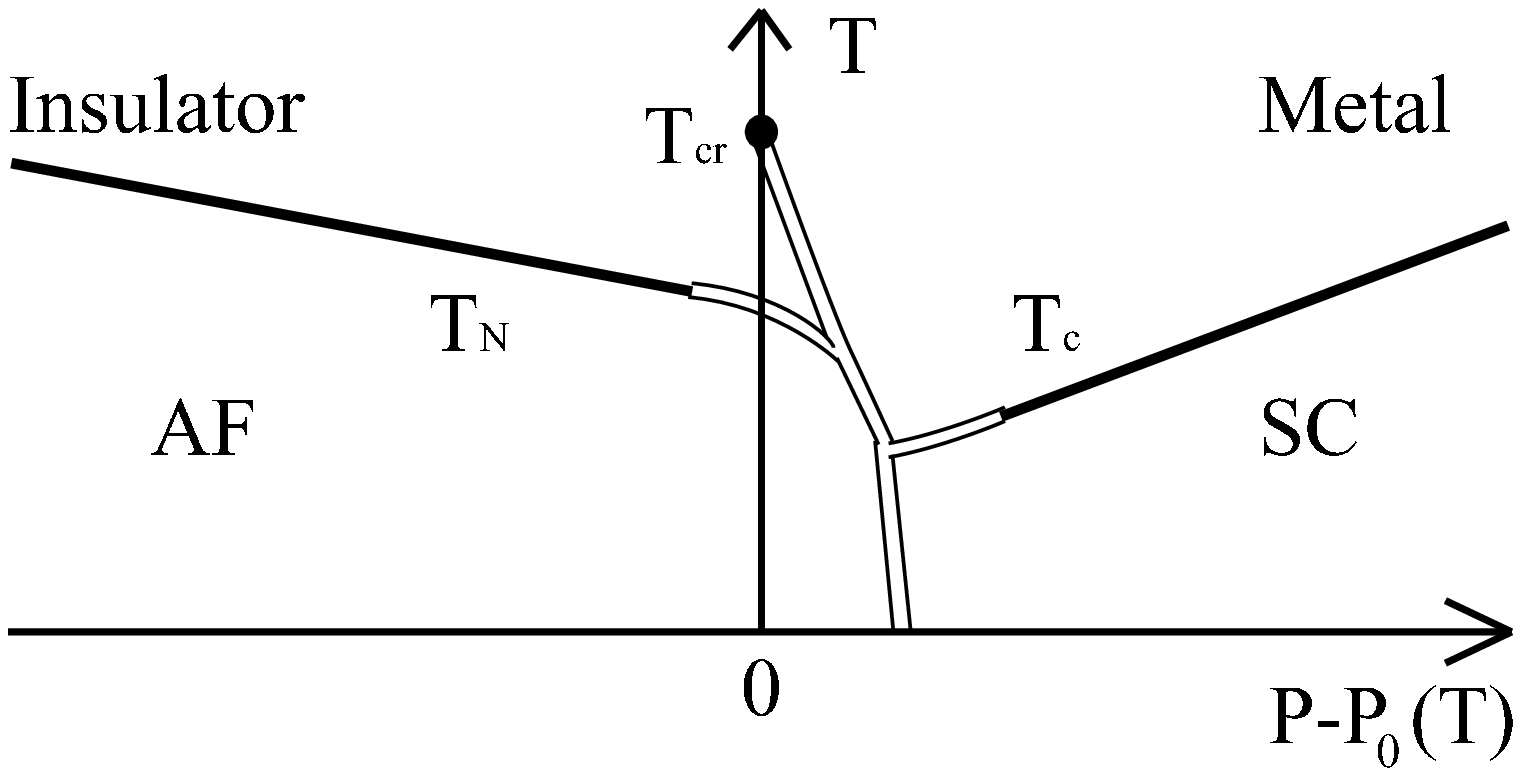}$$
\end{center}
\vspace*{-12mm}
(b)
\caption{RG phase diagrams modified by the fluctuations from the MF one given by Fig.~\ref{fig:MF} (b).}
\label{fig:RG}
\end{figure}

Next, we turn to the RG analysis of the fluctuations. 
We start with the case of Fig.~\ref{fig:MF}(b),
which is crucially modified by the fluctuation into Fig.~\ref{fig:RG}
with the emergence of fluctuation-induced first-order phase transitions. 
The detailed discussion follows below.
Introducing a mometum cutoff $\Lambda=e^\lambda$,
$\tilde{g_i}\equiv\sqrt{I_d}g_i\Lambda^{-\varepsilon_3}$ and $\tilde{u}_{ij}\equiv I_d u_{ij} \Lambda^{-\varepsilon}$ with the integral $I_d=2[(4\pi)^{d/2}\Gamma(d/2)]^{-1}$ with $\varepsilon_3=3-\frac{d}{2}$ and $\varepsilon=4-d$
($d$: the spatial dimension),
we obtain the following  RG equations for the dimensionless running coupling constants up to one-loop order;
\begin{eqnarray}
\tilde{g}_\Delta'&=&-\varepsilon_3 \tilde{g}_\Delta
\!+\!\left[(N_\Delta\!+\!2)\tilde{g}_\Delta \tilde{u}_{\Delta\Delta}\!-\!N_m\tilde{g}_m\tilde{u}_{\Delta m}\right]
\!-\!\tilde{g}_\Delta^3
\\
\tilde{g}_m'&=&-\varepsilon_3 \tilde{g}_m
\!+\!\left[(N_m\!+\!2)\tilde{g}_m\tilde{u}_{mm}\!-\!N_\Delta \tilde{g}_\Delta \tilde{u}_{\Delta m}\right]
\!-\!\tilde{g}_m^3
\\
\tilde{u}_{\eta\eta}'&=&-\varepsilon \tilde{u}_{\eta\eta}
+9\tilde{u}_{\eta\eta}^2+\frac{1}{2}(\tilde{g}_\Delta^4+\tilde{g}_m^4)
\\
\tilde{u}_{\Delta\Delta}'&=&-\varepsilon \tilde{u}_{\Delta\Delta}\!
+\!\left[(\!N_\Delta\!+\!8)\tilde{u}_{\Delta\Delta}^2\!+\!N_m \tilde{u}_{\Delta m}^2\right]\!+\!\tilde{g}_\Delta^4
\\
\tilde{u}_{mm}'&=&-\varepsilon \tilde{u}_{mm}\!
+\!\left[(\!N_m\!+\!8)\tilde{u}_{mm}^2\!+\!N_\Delta \tilde{u}_{\Delta m}^2\right]\!+\!\tilde{g}_m^4
\\
\tilde{u}_{\Delta m}'&=&-\varepsilon \tilde{u}_{\Delta m}+\tilde{g}_\Delta^2 \tilde{g}_m^2
\nonumber\\
&&
+[(\!N_\Delta\!\!+\!2)\tilde{u}_{\Delta\Delta}\!+\!(\!N_m\!\!+\!2)\tilde{u}_{mm}\!+\!4\tilde{u}_{\Delta m}]\tilde{u}_{\Delta m},
\end{eqnarray}
where $N_\Delta=2$ ($N_m =3$) is the dimension of $\vec{\Delta}$ ($\vec{m}$) 
and $'$ represents a derivative with respect to $\lambda$. 

When $g_\Delta=g_m=0$, six fixed points were obtained in terms of $\varepsilon$ expansion \cite{NelsonKosterlitzFisher74}. In particular, there exist two nontrivial fixed points with nonzero $\tilde{u}$'s \cite{MurakamiNagaosa00}: (A) O(N)-symmetric fixed point $\tilde{u}_{\Delta\Delta}=\tilde{u}_{mm}=\tilde{u}_{\Delta m}=\tilde{u}^*_{\rm iso}$ with $\tilde{u}^*_{\rm iso}=\frac{\varepsilon}{N+8}$ and $N=N_\Delta\!+\!N_m$. (B) Biconical fixed point $(\tilde{u}^*_{\Delta\Delta},\tilde{u}^*_{mm},\tilde{u}^*_{\Delta m})\approx(0.09048,0.08475,0.05358)$ for $d=3$. In our case of $N_\Delta=2$, $N_m=3$ and $d=3$, the fixed point (A) is unstable except in a special case $\tilde{u}_{\Delta\Delta}\tilde{u}_{mm}\sim\tilde{u}_{\Delta m}^2$, while (B) is stable.

\begin{figure}[tb]
\begin{center}
\epsfxsize=5.0cm
$$\epsffile{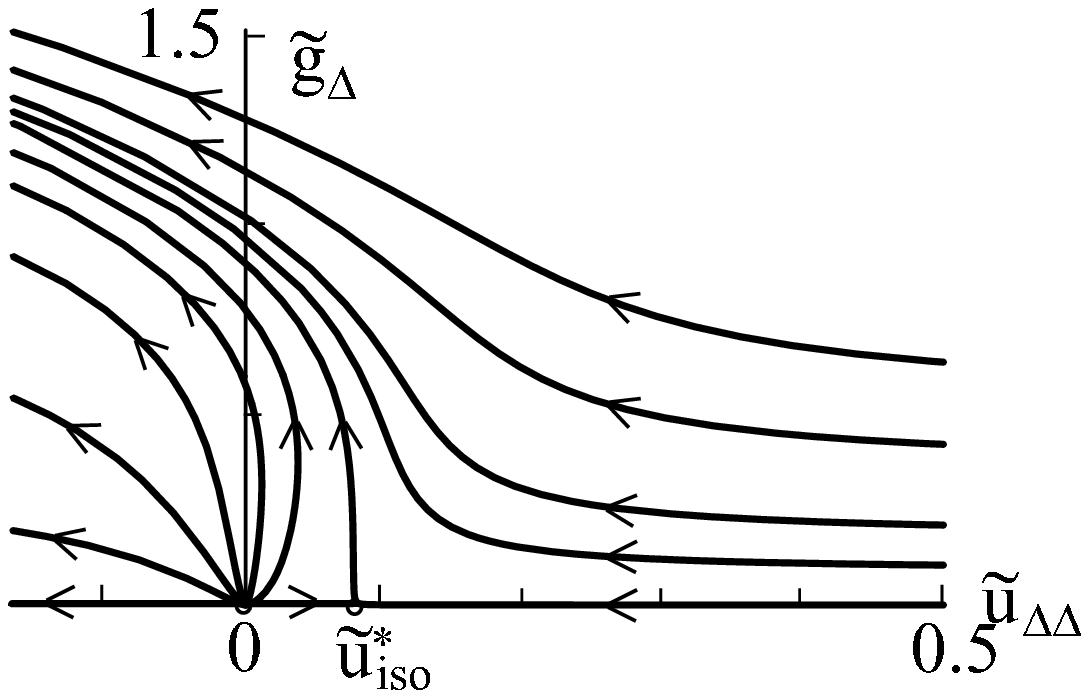}$$
\end{center}
\vspace*{-12mm}
(a)
\begin{center}
\epsfxsize=5.0cm
$$\epsffile{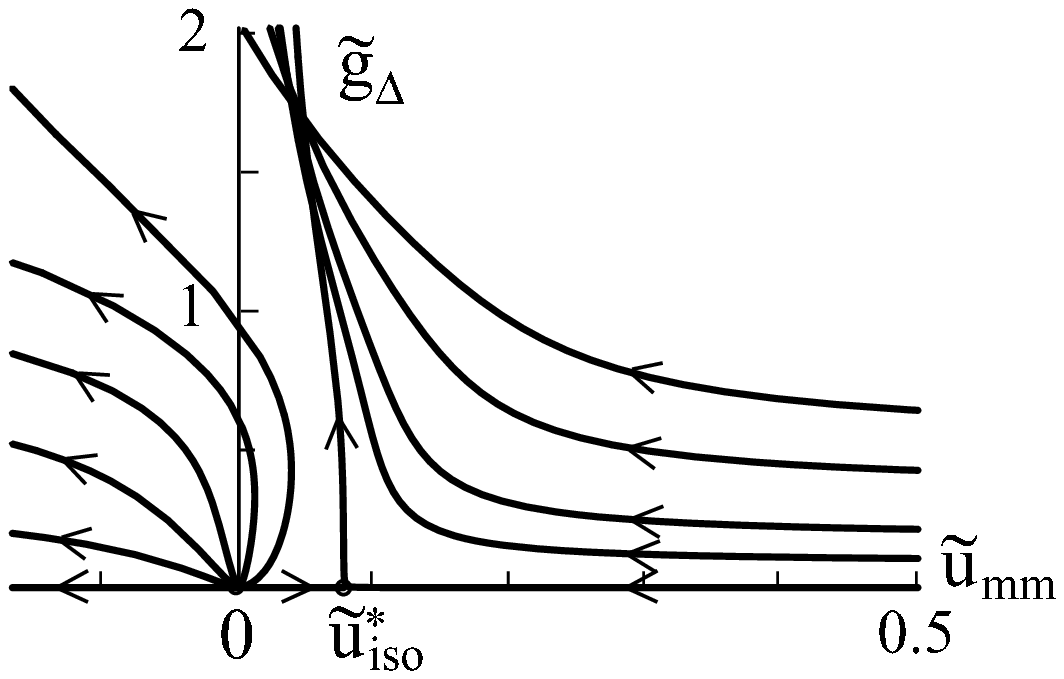}$$
\end{center}
\vspace*{-12mm}
(b)
\begin{center}
\epsfxsize=5.0cm
$$\epsffile{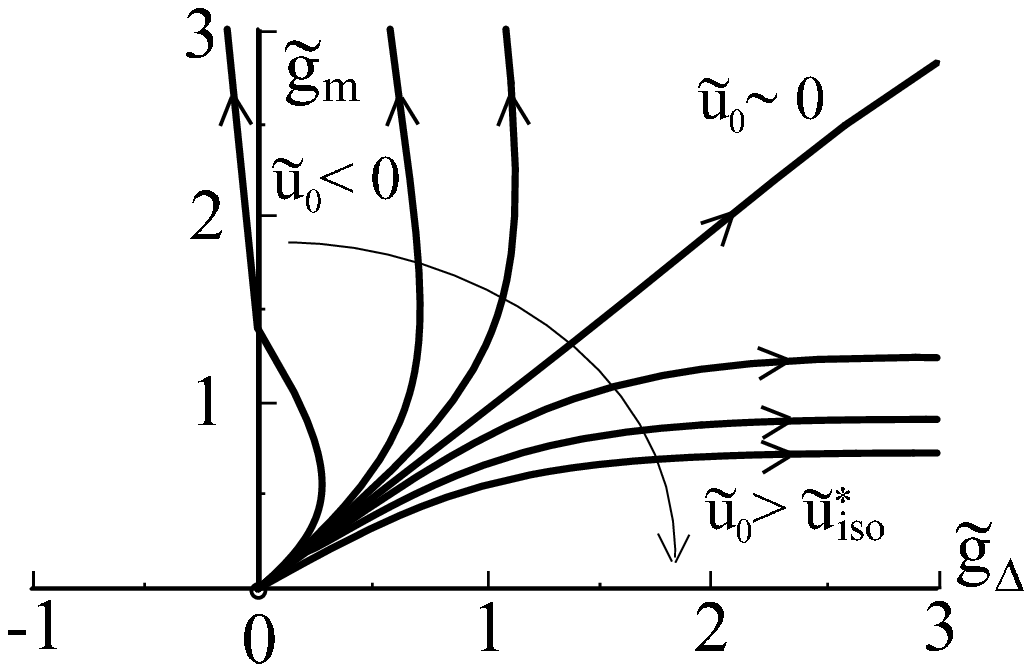}$$
\end{center}
\vspace*{-12mm}
(c)
\caption{RG trajectories around the SO(5)-symmetric fixed point obtained under the initial condition $\tilde{u}_{\Delta\Delta}=\tilde{u}_{mm}=\tilde{u}_{\Delta m}=\tilde{u}_0$ at the scale $\tilde{g}_\Delta$=$\tilde{g}_m\to0$. Curves in (b) correspond to different values of $\tilde{u}_0$.}
\label{fig:flow_O5}
\end{figure}
With $g_\Delta, g_m>0$, however, there exist no nontrivial stable fixed points, i.e.,  $g_\Delta$ and $g_m$ are relevant couplings at the SO(5)-symmetric 
fixed point. 
Figure~\ref{fig:flow_O5} shows the RG trajectories of running coupling constants 
with the SO(5)-symmteric initial condition $\tilde{u}_{\Delta\Delta}=\tilde{u}_{mm}=\tilde{u}_{\Delta m}(=\tilde{u}_0)$ with infinitestimal 
$\tilde{g}_\Delta$=$\tilde{g}_m$. 
$\tilde{u}$'s stay around $\tilde{u}^*_{\rm iso}$ 
as long as $\tilde{g}_\Delta,\tilde{g}_m \lsim 0.5$. 
($\tilde{u}_{\Delta m}$ is not shown in Fig.~\ref{fig:flow_O5}, 
but it behaves similarly to $\tilde{u}_{mm}$.) 
Therefore it is concluded that 
as long as $\tilde{g}_\Delta,\tilde{g}_m \lsim 0.5$, 
the SO(5)-scaling behavior is observed near the ``bicritical point''.
When $\tilde{g}_\Delta,\tilde{g}_m \gsim 0.5$, the instability of the SO(5)-symmetric fixed point is amplified and the bicritical scaling stops. $\tilde{u}$'s flow to $-\infty$, suggesting the fluctuation-induced first-order transitions \cite{fluctuation-induced-1st-order}. The difference among $\tilde{u}$'s rapidly increases in the run-away trajectories far from the SO(5)-symmetric fixed point (Figs.~\ref{fig:flow_O5}(a) and \ref{fig:flow_O5}(b)), because of the difference between $\tilde{g}$'s. Trajectories $(\tilde{g}_\Delta,\tilde{g}_m)$ shown in Fig.~\ref{fig:flow_O5}(c) depend on the initial value of $\tilde{u}_0 =\tilde{u}_{\Delta\Delta}=\tilde{u}_{mm}=\tilde{u}_{\Delta m}$. 
For the choices of $\tilde{u}_0>0$, $\tilde{g}_\Delta$ becomes larger than $\tilde{g}_m$ due to the difference between $N_\Delta=2$ and $N_m=3$, and $\tilde{u}_{\Delta\Delta}$ becomes negative faster than $\tilde{u}_{mm}$ and $\tilde{u}_{\Delta m}$.
Flows of $\tilde{u}_{\eta\eta}$ are also similar to $\tilde{u}_{\Delta\Delta}$, indicating a first-order Mott transition.
These RG results suggest a phase diagram shown in Fig.~\ref{fig:RG}(a) 
with the dominant SC over the AF.
So far we have considered the RG flows around the SO(5) fixed point. 
On the other hand, if we consider the coupling constants far from
the SO(5) fixed point and take $\tilde{g}_m \gg \tilde{g}_\Delta$, 
$\tilde{g}_m$ remains to be larger than $\tilde{g}_\Delta$ and $\tilde{u}_{mm}$ becomes negative faster than $\tilde{u}_{\Delta\Delta}$ and $\tilde{u}_{\Delta m}$. Then, there occurs another possibility of phase diagram shown in Fig.~\ref{fig:RG}(b) with the dominant AF over the SC.
A phase diagram with a fluctuation-induced first-order phase transition to a
coexistent phase is also possible.
To determine which ordered phase appears, we must include higher-order couplings into the free energy and examine the parameters in detail. This is beyond the scope of this paper. 

We turn to the cases of MF phase diagrams, Fig.~\ref{fig:MF}(a) and \ref{fig:MF}(c).
In these cases, the $\eta$ field is massive around $(T_{\rm bcr}, P_{\rm bcr})$, i.e., $|r_\eta|/\Lambda_0^2 \ne 0$ ($\Lambda_0$ is the initial momentum 
cuttoff). 
(I) When $|r_\eta|/\Lambda_0^2 < 1$, the fluctuation of $\eta$ is
relevant and the RG analysis based on eqs.~(1)-(6) applies
for $ \Lambda_d = \sqrt{|r_\eta|}< \Lambda < \Lambda_0$. 
If an instability with negative $\tilde{u}$'s occurs before this scaling stops at $\Lambda_d$, then the phase diagram is again given by Fig.~3(a) or 3(b).
Otherwize, the scaling for $\Lambda<\Lambda_d$ crosses over to that in the following case (II).
(II) When $|r_\eta|/\Lambda_0^2 > 1$, we can neglect the fluctuations of $\eta$,
and the problem is reduced to that of $\vec{\Delta}$ and $\vec{m}$ only
by putting $\eta= \pm \sqrt{ |r_\eta|/ u_{\eta \eta} }$ in the case of 
Fig.~\ref{fig:MF}(a) and  $\eta=0$ in the case of Fig.~\ref{fig:MF}(c).
Furthermore, in the case of Fig.~\ref{fig:MF}(a), the discontinuity $\delta r_{\Delta-m}$ of $r_\Delta-r_m$ at the first-order Mott transition produces another scale for a crossover: The RG flows for $\tilde{u}_{\Delta \Delta}$, $\tilde{u}_{m m}$, and $\tilde{u}_{\Delta m}$ with $\tilde{g}$'s being zero 
in eqs.~(4)-(6) \cite{MurakamiNagaosa00} should be stopped at 
$\Lambda_c = \sqrt{\delta r_{\Delta - m}}$.
Therefore, when $\delta r_{\Delta - m} > \Lambda_0^2$ in Fig.~\ref{fig:MF}(a), 
AF and SC are well separated by the first-order Mott transition, while 
the scaling theory in ref.~16 applies 
for the range of $\Lambda_c<\Lambda< \Lambda_0$ in the case of 
$\delta r_{\Delta - m} < \Lambda_0^2$.  

 Now we discuss the phase diagram of $\kappa$-(BEDT-TTF)$_2$X shown in 
Fig.~\ref{fig:exp}.
The Mott critical point is well above $T_{\rm bcr}$, and 
the phase diagram looks like Fig.~\ref{fig:MF} (a). 
Therefore SC and AF occur independently separated by the first order
Mott transition, and do not interact with each other. 
However $T_{\rm N}$ and $T_{\rm c}$ almost meet (within 1 K) 
at the ``bicritical point'', and the observed singular behavior of 
NMR relaxation rate strongly suggests the SO(5) bicritical scaling
\cite{MurakamiNagaosa00}. These suggest that SC and AF communicate with 
each other across the Mott transition. This dilenma could be resolved
by assuming small $\delta r_{\Delta-m}^{} / \Lambda_0^2$, because 
the bicritical scaling governed by the SO(5) fixed point should be observed 
in a certain region before the coupling to $\eta$ grows. 
This small value of $\delta r_{\Delta-m}^{} / \Lambda_0^2$ also leads to fluctuation-induced first-order phase transitions 
to SC and/or AF, as obtained in the RG analysis (Fig.~\ref{fig:RG}). 
This is not excluded in Fig.~\ref{fig:exp}, 
because the first-order Mott transition must be accompanied by 
a hysteresis region which defines metastabilities of metallic and insulating 
states. The fluctuation-induced first-order phase transitions to SC and/or AF 
might be hidden in this hysteresis region, and further experiments on the 
metastability are required. Moreover, if $T_{\rm cr}$ approaches $T_{\rm bcr}$ 
in other materials by varying the geometrical frustration 
\cite{OnodaImada03lett}, then the present scenario has a more chance to be 
observed experimentally.

Lastly, we discuss critical properties around the Mott critical point $(P_{\rm cr}, T_{\rm cr})$. The critical properties are described by the Ising universality class. Then, the specific heat, the derivative of $\langle \eta \rangle$ with respect to $h$ and its jump $\delta \langle \eta \rangle$ obey the following
scaling laws, 
\begin{eqnarray}
C\propto|T-T_{\rm cr}|^{-\alpha}
\\
\partial \langle \eta \rangle / \partial h |_{T,\ h=h_{\rm cr}} &\propto& (T-T_{\rm cr})^{-\gamma}
\mbox{ for $T>T_{\rm cr}$},
\\
\delta \langle \eta \rangle |_{T,\ h=h_{\rm cr}} &\propto& (T_{\rm cr}-T)^\beta
\hspace*{2mm}\mbox{ for $T<T_{\rm cr}$},
\\
\delta \langle \eta \rangle |_{T=T_{\rm cr},\ h} &\propto& |h-h_{\rm cr}|^{1/\delta}
\end{eqnarray}
with $\alpha\sim0.013$, $\gamma\sim1.25$, $\beta\sim0.31$ and $\delta\sim 5.0$ in three dimensions \cite{Ma}. Here, $h_{\rm cr}$ is the critical value of $h$ and $h_{\rm cr}=0$ if we take $\vec{\Delta}=\vec{m}=0$ in our free energy. We note that the double occupancy $D$ and $U$ play roles of density and pressure, respectively, in the analogy to the liquid-gas phase transition \cite{Castellani79,DMFT,OnodaImada03lett}. Therefore, $D$ and $\partial D / \partial U$ exhibit the same critical properties as $\langle \eta \rangle$ and $\partial \langle \eta \rangle / \partial h$, and hence the local magnetic moment given by $\frac{1}{4}-\frac{D}{2}$ in the Hubbard model.
The singularity at the critical point and the discontinuity at the first-order phase boundary can also be observed in other physical quantities such as resistivity, Drude weight, spin susceptibility, and longitudinal and transverse thermal conductivities. This is because these quantities can be expanded in terms of $\langle \eta \rangle / T$ and the first-order term exhibits a dominant singularity. Then, in the vicinity of the Mott critical point, the same scaling properties as $\eta$ can be observed by replacing $h$ with $P$. Especially, the thermal Hall conductivity may be useful for the experimental scaling analysis, since phonon contributions are usually negligible.

We have considered only the half-filling case, but one can consider the electron filling $n$ as another controlling parameter in addition to $U/W$. In particular, the divergence of $\kappa_{\rm el}$ at the critical point \cite{KotliarMurthyRozenberg02,OnodaImada01full} has been observed as an anomaly of the sound velocity $c$ in ultrasonic measurements by Fournier {\it et al.} \cite{Fournier03}, although it is still difficult to estimate the scaling property from the experiment. The same anomaly for $c$ can also be obtained with Raman scattering experiments. Extention of the present GL theory to include $n$ is an interesting problem left for future studies.

The authors acknowledge K. Kanoda, K. Miyagawa and F. Kagawa for stimulating discussion and providing experimental data before publication. They also thank S. Murakami for fruitful discussion.

\end{document}